\documentclass[reprint,longbibliography]{revtex4-1}
\setcitestyle{super,open={},close={}}
\usepackage[utf8]{inputenc}
\usepackage{amsmath}
\usepackage{amsfonts}
\usepackage{amssymb}
\usepackage{graphicx}

\usepackage{color}
\usepackage[mathlines]{lineno}

\begin{document}
\title{Transition metal impurities in Silicon: Computational search for a semiconductor qubit}

\author{Cheng-Wei Lee}
\affiliation{Colorado School of Mines, Golden, CO 80401, USA}
\affiliation{National Renewable Energy Laboratory, Golden, CO, 80401, USA}
\author{Meenakshi Singh}
\affiliation{Colorado School of Mines, Golden, CO 80401, USA}
\author{Adele Tamboli}
\affiliation{Colorado School of Mines, Golden, CO 80401, USA}
\affiliation{National Renewable Energy Laboratory, Golden, CO, 80401, USA}
\author{Vladan Stevanovi\'c}
\email{vstevano@mines.edu}
\affiliation{Colorado School of Mines, Golden, CO 80401, USA}
\affiliation{National Renewable Energy Laboratory, Golden, CO, 80401, USA}

\begin{abstract}
Semiconductors offer a promising platform for the physical implementation of qubits, demonstrated by the successes in 
quantum sensing, computing, and communication. The broad adoption of semiconductor qubits is presently hindered by 
limited scalability and/or very low operating temperatures. 
Learning from the NV$^{-}$ centers in diamond, whose 
optical properties enable quantum sensing at room temperature and quantum communication at low temperature, our goal is to find equivalent optically active point defect centers in crystalline silicon, which could be advantageous for their scalability and integration with classical devices. 
Motivated by the fact that transition metal (TM) impurities in silicon are typically paramagnetic deep defects, we apply generalized density functional theory (HSE06) methods to investigate electronic and optical properties of these deep-level defects.
Standard HSE06 is known to overlocalize $d$ orbital of TM and the error can be estimated by non-Koopmans' energy(E$\mathrm{_{NK}}$), which characterizes the deviation from generalized Koopmans' condition (gKC). 
When the deviation is too large for the searching purpose, we follow the established correction scheme that utilizes an occupation-dependent potential, which is determined self-consistently by satisfying gKC, on $d$ orbital of TM.
A comprehensive investigation for the whole 3d transition metals in Si at this theory level is missing in the literature and we use the newly acquired results to examine their potential as optically active spin qubits in Si.
We identify seven transition metal impurities that have optically allowed triplet-triplet transitions within the Si band gap, which could be considered candidates for such qubits. 
These results provide the first step toward Si-based qubits with higher operating temperatures for quantum sensing.  Furthermore, spin-photon interfaces in Si have potential application for coupling and readouts of spin donor qubits in Si and for mid-infrared free-space communications. 
\end{abstract}

\maketitle 

\section*{\label{sec:intro}Introduction}
Quantum information science has drawn a lot of attention in the past few decades due to its potential to transform how information is collected, processed, and transmitted\cite{NSF_QIS_1999,NSTC_SCQIS_2018,Acin_NJP_2018}. 
Recent developments focus on the physical realization of qubits\cite{Ferrenti_npj_CM_2020}, in which quantum information is encoded.
Semiconductor qubits are among the promising systems as demonstrated by the successes in quantum sensing\cite{Elzerman_Nature_2004,Liu_PRL_2019}, quantum computing\cite{Plat_nature_2013,Saeedi_science_2013,Veldhorst_nature_nanotech_2014,Abobeih_Nature_Comm_2018}, and quantum communication\cite{Yin_Nature_2013,Hensen_nature_2015}.

To realize a qubit, a quantum two-level system whose levels can be initialized, coherently controlled and measured with high fidelity is needed\cite{Chatterjee_arxiv_2020}.
For semiconductors, three major qubit types are discussed\cite{Chatterjee_arxiv_2020}: ({\it i}) gate-controlled nanostructures, ({\it ii}) shallow dopants in silicon, and ({\it iii}) optically-addressable point defects such as NV$^{-}$ centers in diamond, a negatively charged defect complex between substitutional nitrogen impurity and an adjacent carbon vacancy.
All these qubit types aim to isolate a charge or spin from the semiconductor host and thus form the required quantum two-level systems. 
Their performance is benchmarked against different quantum applications and each has their unique strengths and weaknesses.\cite{Chatterjee_arxiv_2020}

NV$^{-}$ centers in diamond and neutral (and shallow) $^{31}$P impurities in crystalline Si are exemplary systems for optically-addressable point defects and long coherence times, respectively.
The NV$^{-}$ centers have been demonstrated to work even at room temperature, which is critical for quantum sensing\cite{Steinert_RSI_2010}. 
Also, their optical properties offer a natural spin-photon interface suitable for quantum communication at low temperature.
The NV$^{-}$ centers have been shown to achieve quantum communication at a distance over one kilometer (at 4K)\cite{Hensen_nature_2015}, but are ultimately limited by the strong attenuation of their 637-nm light through optical fiber\cite{Chatterjee_arxiv_2020}.
Optically-addressable point defects in general meet the criteria for quantum computing, but the probabilistic nature of where the defects are created limits the scalability, which is essential for quantum computing\cite{Chatterjee_arxiv_2020}. 
$^{31}$P in Si currently has the longest reported decoherence time for an electron spin in a semiconductor qubit ($\approx$ few seconds)\cite{Muhonen_nnano_2014}. 
The even longer decoherence time for nuclear spin\cite{Muhonen_nnano_2014} ($>$10 s) makes such systems relevant for the storage of quantum information, i.e., quantum memories.
However, low temperatures ($<$10 K for $^{31}$P in Si) are required.

Among practical considerations, scalability and operating temperature stand out as the major challenges for semiconductor qubits\cite{Chatterjee_arxiv_2020,Wolfowicz_NRM_2021}.
The mature manufacturing techniques for Si-based devices makes Si an attractive host material but the extremely low operating temperature is a major concern.
Besides, silicon-based implementations currently lack established spin-photon interfaces, which is critical for application in quantum communication\cite{Chatterjee_arxiv_2020, Redjem_nature_electronics_2020, Zhang_APR_2020} and for an alternative scheme to couple and read out spin donor qubits in Si\cite{Morse_SA_2017}.
On the other hand, NV$^{-}$ centers in diamond operate at room temperature, but scaling-up for diamond-based devices remains challenging. 
Considering the advantages of both systems, one potential approach to address the challenge is to find a NV$^{-}$  center like defect system in silicon.
Specifically, resembling NV$^{-}$  centers in diamond, we are searching for deep centers that can accommodate optical triplet-triplet transitions within the Si band gap.

Color centers were rarely observed for silicon due to its small electronic band gap.
A recent review summarizes the known color centers, including $^{77}$Se$^{+}$, Er$^{3+}$, and three color centers associated with particle irradiation\cite{Zhang_APR_2020}.
Among these color centers, T- and G-centers in Si were recently reported to have in-gap optical transition at the O-band of optical telecommunications\cite{Redjem_nature_electronics_2020,Bergeron_PRXQuantum_2020}. 
Redjem \emph{et al.} demonstrated a single photon emitter in silicon for the first time using C-doped Si\cite{Redjem_nature_electronics_2020}.
The single photon emitter is currently associated with the G-center and has spin singlet-singlet transition at $\approx$ 1.27 um\cite{Redjem_nature_electronics_2020}.
Similarly, Bergeron \emph{et al.}  reported the spin and optical properties of an ensemble of T-centers in Si, which has long-lived electron spins ($\approx$ ms) and spin-selective transitions at 1326 nm\cite{Bergeron_PRXQuantum_2020}.
However, no color centers in silicon were found for the same triplet-triplet optical transitions as NV$^{-}$ centers in diamonds.

Transition metal (TM) impurities in Si have been widely studied computationally and experimentally in the literature due to scientific and technological interests\cite{Weber_APA_1983, Beeler_PRB_1990, Heiser_book_2004}. 
Their electron paramagnetic resonance measurements can be largely explained by the model proposed by Ludwig and Woodbury, which is based on the tetrahedral crystal field and Hund's rule\cite{Weber_APA_1983, Beeler_PRB_1990}. 
Later at 1990, Beeler \emph{et al.} applied density functional theory (with the LDA functional) based Green's function method to investigate the electronic properties of the whole 3d TM impurities in Si except for Zn\cite{Beeler_PRB_1990}. 
They focused on \emph{unrelaxed} TM impurities on tetrahedral interstitial and substitutional sites and their prediction is qualitatively consistent with experimental measurements but deviates from the Ludwig and Woodbury model. 
Beeler \emph{et al.} found that Hund's rule breaks down for the early TM interstitial and the late TM substitutional impurities in Si.
Their work is the first comprehensive first-principle studies on TM  impurities in Si and TM impurities with spin triplet ground state were readily identified. 
However, their work is limited by the accuracy of LDA regarding Si band edges and the lack of structural relaxation, which together could give rise to qualitatively different predictions. 
Generalized density functional theory using HSE06 method can accurately predict Si band gap but it also overlocalizes $d$ orbital of TM in Si  due to its homogeneous screening\cite{Ivady_PRB_2013}.
Multiple approaches can address this issue\cite{Ivady_PRB_2013,Ivady_PRB_2014,Zheng_PRMaterials_2019} and one of them was demonstrated on Fe interstitial defects in Si\cite{Ivady_PRB_2013}.
But a comprehensive study at such improved theory level, which are needed for the search, is still missing in the literature.
In addition, no optical properties are reported for those TM impurities with triplet ground states and their potentials for optically active spin qubits remain elusive. 

\begin{figure}[!htb]
\centering
\includegraphics[width=0.75\linewidth]{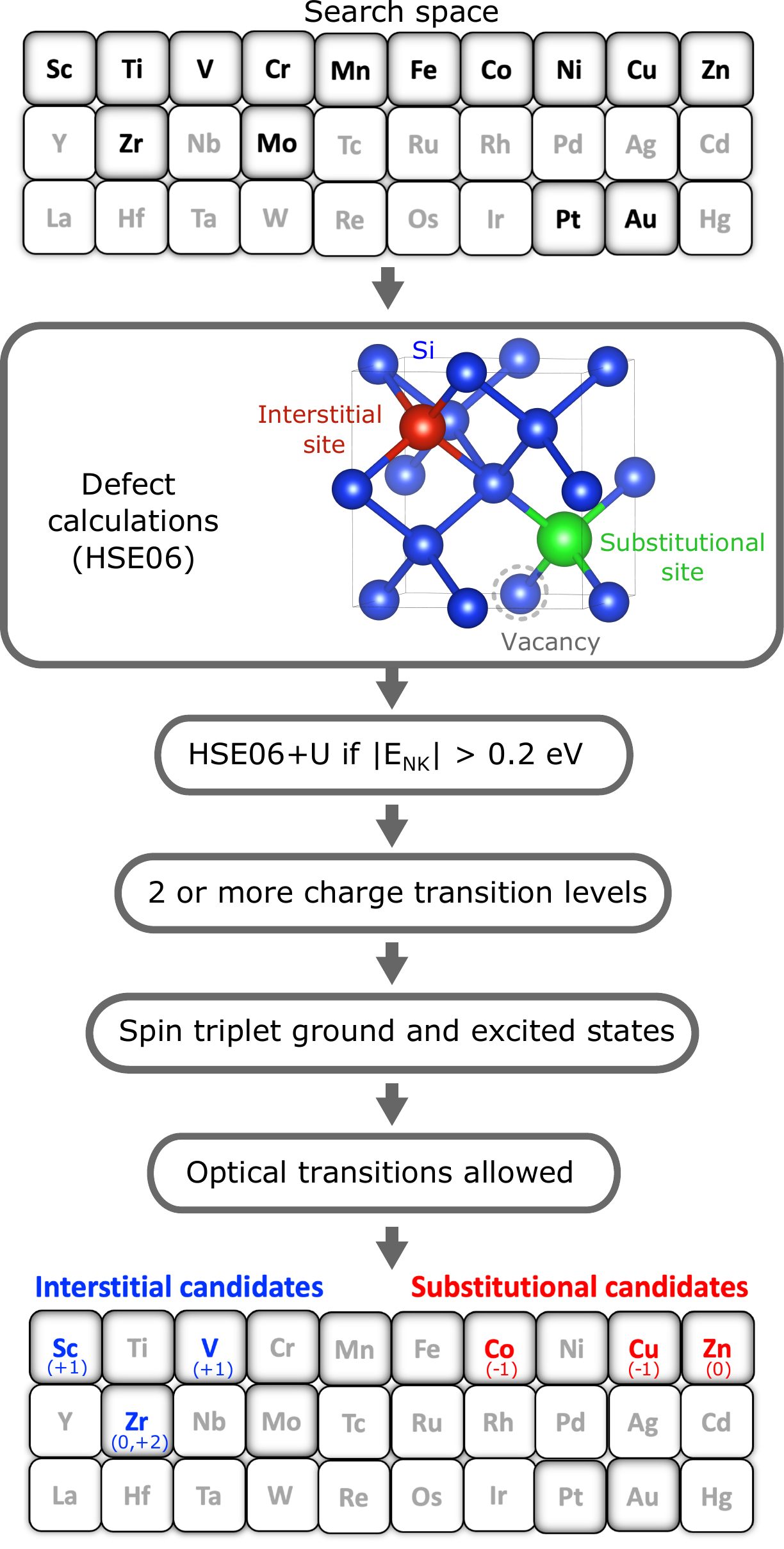}
\caption{\label{fig:PT_trend} 
\textbf{Workflow adopted in our search for transition metal impurities in crystalline Si.} 
The search space consists of all 3d and select 4d and 5d transition metals (shown at the top). 
Defect calculations then enable reducing the set using various search criteria, leading to the identification of 7 candidate impurities as interstitials and substitutional defects (shown at the bottom with specific charge state).  
E$\mathrm{_{NK}}$ is the non-Koopmans' energy and characterizes the deviation from generalized Koopmans' condition.
}
\end{figure}

Motivated by the opportunities and challenges described above, we apply the state-of-art defect calculations based on HSE06, which address the issues of band edge correction and structural relaxation of the previous work.
To address the issue of HSE06 overlocalizing TM, we adopt the approach by  Iv\'ady \emph{et al}.\cite{Ivady_PRB_2013,Ivady_PRB_2014}
The deviation from generalized Koopmans' condition (gKC) is evaluated for the HSE06 results  and the correction term, which is an occupation-dependent potential, is applied on $d$ orbitals of TM when the non-Koopmans' energy is too large for the search ($|$E$\mathrm{_{NK}}|$ $>$ 0.2 eV). 
0.2 eV is chosen in a way to balance computation cost and accuracy since enforcing gKC (E$\mathrm{_{NK}}$ = 0.0 eV) for all the transitions in this work is unfeasible.  
We apply this approach to the whole 3d and few selected heavier TMs, as highlighted in Fig.\ \ref{fig:PT_trend}.
Heavier transition metals are chosen based on the potential for ion implantation\cite{Bischoff_APR_2016, Pacheco_RSI_2017} and literature search on defect levels\cite{Chen_ARMS_1980,Weber_APA_1983}.
Fig.\ \ref{fig:PT_trend} shows the general workflow we developed and adopted to identify promising candidates.
We utilize the present state-of-art computational approach to evaluate defect formation energies, thermodynamic charge transition levels (CTL), and one-particle defect-level diagrams (DLD) for each TM as a substitutional, and an interstitial point defect in crystalline Si, as well as a Si-vacancy/TM-substitutional defect complex. 
Substitutional and interstitial point defects are the most common extrinsic defect types and vacancy complexes are widely found as color centers in other wide band gap host materials such as NV$^{-}$ centers in diamond. 
We also calculate optical absorption spectra to verify that the in-gap optical transitions are allowed. 
Using the selection criteria discussed later in details, we identify three substitutional and four interstitial TM defects in silicon depicted at the bottom of Fig.\ \ref{fig:PT_trend}.
While further investigations into their optical properties and spin dynamics are required, our discovery is the first step toward Si-based qubits with higher operating temperature for quantum sensing and spin-photon interfaces that can couple and read out spin donor qubits in silicon.  
Separately, the optical transition at mid-infrared range can be potentially utilized for free-space communication\cite{Hao_AO_2017,Lin_Nanophotonics_2018}

\section*{\label{sec:results}Results}
In search of candidate TM impurities / Si-host system with triplet-triplet optical transition within the Si band gap, Fig.\ \ref{fig:PT_trend} summarizes our computational workflow.
For every defect, we employ the defect calculations based on HSE06 and first compute the thermodynamic charge transition levels (CTL) that can be readily extracted from the defect formation energies (see Fig. S1 for the explicit data).
E$\mathrm{_{NK}}$ is evaluated for each transition (see Table S6) and further correction (HSE06+U) is applied when $|$E$\mathrm{_{NK}}|$ $>$ 0.2 eV.
For a given impurity defect in Si to accommodate both a spin triplet ground state and a spin triplet excited state within the Si  band gap, at least two CTL within the band gap are needed (see details in METHODS). 
For those impurities that satisfy the first criterion, we further examine their single-particle defect-level diagram (DLD) since CTL have no direct information about the spin configuration and experimental values are only available for a few TM defects Si with specific charge states and defect types. 
DLD shows the ground-state spin configurations for TM impurities and we first search for the ones with spin triplet electronic ground state, i.e., two unpaired electrons in the same spin channel. 
Next, to ensure that a spin triplet excited states exist within the Si band gap, we also search for the TM impurities with at least one empty state within the Si band gap such that one of electrons can be promoted to the empty state(s) and form a spin triplet excited state.
In addition, we use the charge density of corresponding defect states to determine the symmetry of the electronic orbitals and confirm that the single-particle states within the band gap are localized.
Localized spin defect states are required for optically active spin defects\cite{Weber_PNAS_2010} and symmetries provide the foundation to estimate whether the optical transitions are allowed.
Lastly, because the symmetry provides no information about the actual values of the amplitudes (oscillator strengths) of allowed optical transitions, we calculate the one-particle absorption coefficient to verify that the specific triplet-triplet optical transitions are indeed allowed and to estimate their transition probability. 

%
\subsection*{Substitutional Defect Candidates}
%
\begin{figure}[!bht]
\centering
\hspace*{-0.2cm}\includegraphics[width=1.05\columnwidth]{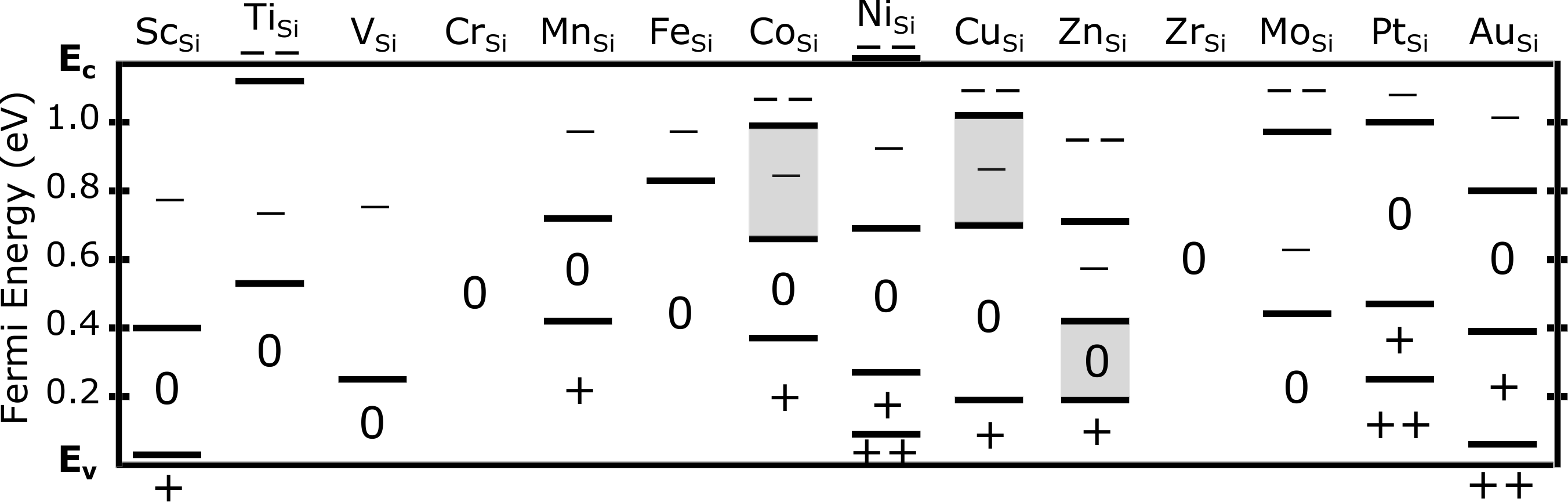}
\caption{\label{fig:CTL_sub} \textbf{Thermodynamics charge transition levels for substitutional defects of transition metals.}
Semi-transparent gray areas highlight the Fermi-level range for the final candidates with triplet-triplet optical transitions within the Si band gap to be stable. 
$0$, $+$, and $-$ signs indicate the dominant charge state within the Fermi-level range.
}
\end{figure}
We first look at the substitutional defects for the transition metals in silicon.
Fig.\ \ref{fig:CTL_sub} shows their thermodynamic charge transition levels within the silicon band gap. 
Most CTL are based on HSE06 calculations except for those of Ti$\mathrm{_{Si}}$ and Mo$\mathrm{_{Si}}$, which are corrected due to large E$\mathrm{_{NK}}$. 
Our HSE06(+U) results generally agree well with experimental results (see METHODS) for substitutional TM defects in Si that have established CTL within the uncertainty of band edge ($\approx$0.1 eV).
Two exceptions are the (-1/-2) level for Co$\mathrm{_{Si}}$ and (+1/0) level for Zn$\mathrm{_{Si}}$, to which no experimental deep-level transient spectroscopy peaks are assigned. 
Since such difference not necessarily disapproves our prediction, further investigation is needed to resolve the difference.
Nonetheless, excluding these two levels has no effect on the screening results since the charge states in question (Co$\mathrm{_{Si}^{-2}}$  and Zn$\mathrm{_{Si}^{}+1}$) have no stable spin triplet ground state.
For those less studied substitutional TM defects in Si, our HSE06(+U)  results provide theoretical reference for future experimental investigation. Detailed comparison for each TM can be found in SM.
Applying the the criteria of at least two CTL excludes V$\mathrm{_{Si}}$, Cr$\mathrm{_{Si}}$, Fe$\mathrm{_{Si}}$, and Zr$\mathrm{_{Si}}$.
%
\begin{figure}[!hbt]
\centering
\includegraphics[width=0.85\linewidth]{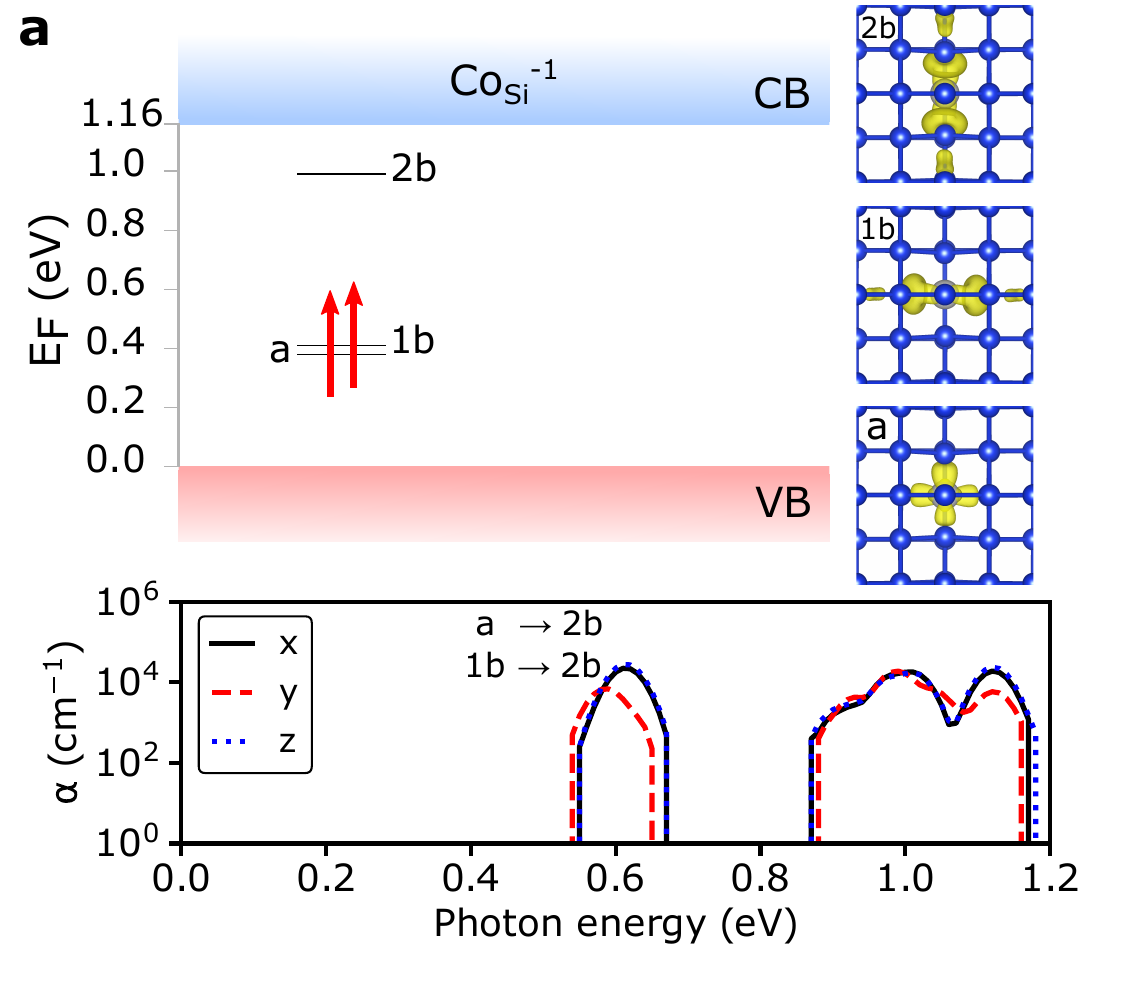}
\includegraphics[width=0.85\linewidth]{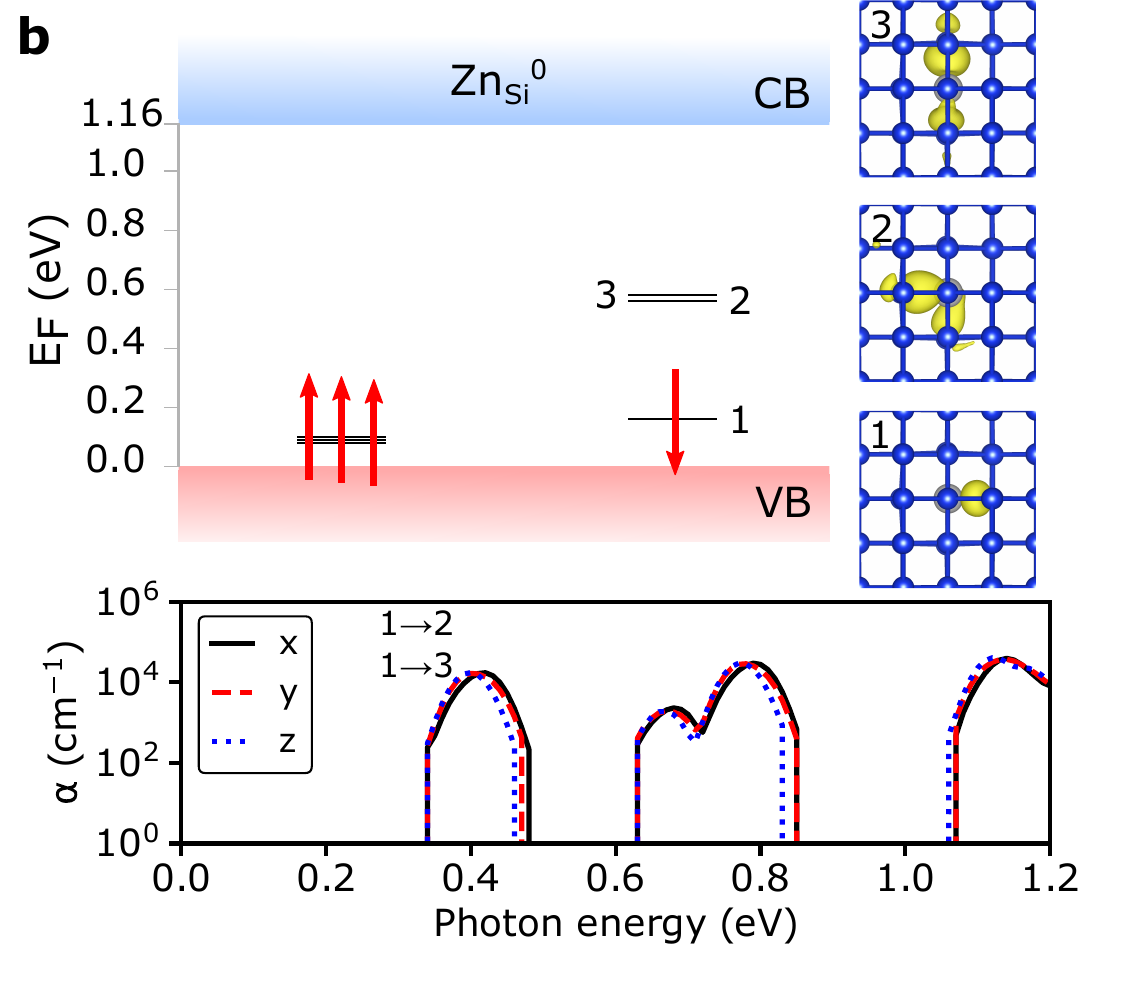}
\caption{\label{fig:sub_candidate} 
\textbf{Single-particle defect level diagram and absorption spectra for representative substitutional defects of transition metals.}
(a) Co$\mathrm{_{Si}^{-1}}$ has C$_{2}$ symmetry while (b) Zn$\mathrm{_{Si}^{0}}$ has no symmetry. 
The numbers are used to distinguish different defect states while letters highlight their symmetries.
The isosurface plots of the charge density of corresponding defect states are shown at the density of 0.0025 $\frac{1}{a_{B}^{3}}$.
Absorption coefficient ($\alpha$) are provided, with specific peaks indicated for the corresponding transitions between labeled defect states within the band gap.
x, y, and z indicate the polarization direction. 
All the other substitutional defect candidates (Cu$\mathrm{_{Si}^{-1}}$ ), which also have in-gap triplet-triplet optical transitions,  can be found in the Supplemental Materials.
}
\end{figure}

Next, we search for the defect systems with stable spin triplet ground states, i.e., having stable charge state within the band gap and spin state of $S$=1 (see Supplemental Materials for the predicted total spin quantum numbers of all the substitutional defects).  For this purpose, we analyze the single-particle defect-level diagrams (see Fig.\ \ref{fig:sub_candidate}) in addition to CTL.
This criterion significantly narrows the list to Sc$\mathrm{_{Si}^{+1}}$, Mn$\mathrm{_{Si}^{+1}}$, Co$\mathrm{_{Si}^{-1}}$,  Ni$\mathrm{_{Si}^{-2}}$, Cu$\mathrm{_{Si}^{-1}}$, Zn$\mathrm{_{Si}^{0}}$, and Mo$\mathrm{_{Si}^{0}}$.
Ni$\mathrm{_{Si}^{-2}}$ is also included since it is within the uncertainty of the band edges. 
We note that Cu$\mathrm{_{Si}^{-1}}$ prefers $S$=1 over $S$=0 by ~13 meV, which is qualitatively different from a previous HSE06 study using 64-atom cell\cite{Sharan_PRAppl_2017}.
The difference is most likely due to cell size effect since significant dispersion of defect levels are observed for Si 64-atom supercell in the literature\cite{Goyal_CMS_2017}.
Mo$\mathrm{_{Si}^{0}}$ and Mn$\mathrm{_{Si}^{+1}}$, though have stable $S$ = 1 configuration, are dropped from the list since the associated defect levels are within the Si valence band. 
Since triplet-triplet optical transitions are desired, we also used the single-particle diagram to determine if there are any spin triplet excited states within the band gap. 
Sc$\mathrm{_{Si}^{+1}}$  and Ni$\mathrm{_{Si}^{-2}}$ are  further disqualified since they have no spin-triplet excited states available within the band gap.

Fig.\ \ref{fig:sub_candidate} shows the electronic single-particle levels diagram for two example substitutional defects out of three candidates, Co$\mathrm{_{Si}^{-1}}$ and Zn$\mathrm{_{Si}^{0}}$ (see Fig.\ S2 in SM for all the substitutional defect candidates).
Fig.\ \ref{fig:sub_candidate}(a) and (b) clearly show the spin triplet ground state with either two up spins or three up spins compensated by one down spin, respectively.
The spin triplet excited states can be achieved when one of the up or down spins was excited under spin-conservation transition for the case of two up spins or three up spins compensated by one down spin, respectively.
Besides, we also remove candidates with defect levels that are too close to the band edge ($\Delta E$ $<$ 0.05 eV) since electrons on these defect levels can be easily thermally excited. The optical transitions between defect levels and band edges also compete with transition between in-gap defect levels, giving rise to optical signals that are too close to distinguish. 
We further confirm the localization of those mid-gap defect levels by visualizing their charge density (see Fig.\ \ref{fig:sub_candidate}). 
These charge densities also support the symmetry determination for the defect based on atomic configurations. 
For the candidates without symmetry, e.g. Fig.\ \ref{fig:sub_candidate}(b), we can clearly see the distortion from mirror or rotation symmetries.

Lastly, we calculate the one-particle optical absorption coefficient for these five candidates (see bottom subfigures in Fig.\ \ref{fig:sub_candidate}(a) and (b)), in order to verify that the optical triplet-triplet transitions are not only symmetry allowed but also have high transition probability. 
We focus on the photon energy range within the silicon band gap since we are searching for in-gap optical transitions.
The first strong peaks are related to the triplet-triplet transitions for all the candidates while the rest of the peaks are associated with transition between band edge states and the mid-gap defect states. 
Also, the absorption coefficient for all first peaks reaches $\sim$10$^{3}$-10$^{4}$ cm$^{-1}$, suggesting relatively high transition probability. 
With all these criteria, we found three promising substitutional TM defects in Si (Co$\mathrm{_{Si}^{-1}}$, Cu$\mathrm{_{Si}^{-1}}$, and Zn$\mathrm{_{Si}^{0}}$).
We note that their in-gap triplet-triplet transitions are within the mid-IR photon energy range of 0.4 - 0.6 eV ($\approx$ 3100 - 2066 nm).
Such mid-IR photons have significantly larger attenuation loss (at the order of 10-1000 dB/km) through silica optical fiber\cite{Artyushenko_OP_2014}, in comparison to the NV$^{-}$ center (1-10 dB/km) and the optimal telecommunication bands (few tenth dB/km)\cite{Artyushenko_OP_2014,Wolfowicz_NRM_2021}, making them unsuitable for quantum communication using silica optical fiber. 
But mid-IR photons fall in the so-called atmospheric windows\cite{Hao_AO_2017}, which are the desired wavelength for free-space long-distance communication. 
Since localized defects are known to provide an alternative way to emit mid-IR radiation\cite{Lin_Nanophotonics_2018}, the candidates identified here have the potential to advance free-space long-distance communication. 
For quantum applications, our results reveal the candidates that can potentially serve as spin-photon interfaces, which can couple and read out spin donor qubits in Si\cite{Morse_SA_2017}.
Furthermore,  the recently reported G- and T-centers in Si\cite{Redjem_nature_electronics_2020,Bergeron_PRXQuantum_2020} fall in this telecommunication range and showcase the possibility to use Si as platform for quantum communication at low temperature (few K).
Therefore, for quantum communication, future search for color centers with telecommunication wavelengths in Si remains open.

\subsection*{Interstitial Defect Candidates}
\begin{figure}[!htb]
\centering
\hspace*{-0.3cm}\includegraphics[width=1.05\columnwidth]{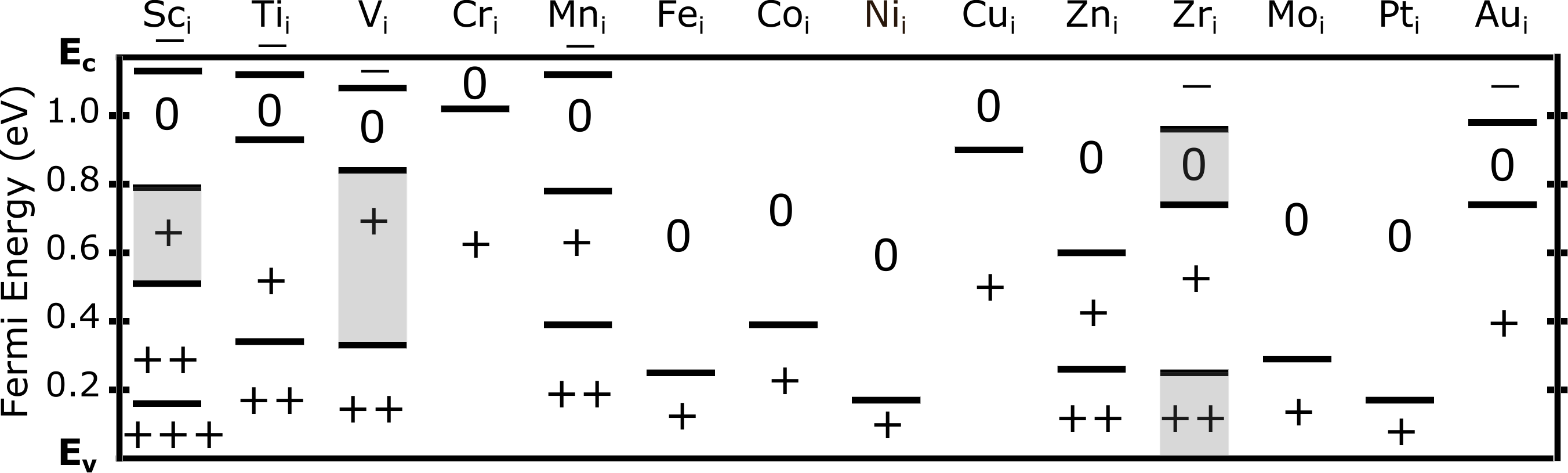}
\caption{\label{fig:CTL_int} \textbf{Thermodynamics charge transition levels for tetrahedral interstitial defects of transition metals. }
Semi-transparent gray areas highlight the Fermi-level range for the final candidates with triplet-triplet optical transitions within the Si band gap to be stable. 
$0$, $+$, and $-$ signs indicate the dominant charge state within the Fermi-level range.
}
\end{figure}

\begin{figure}[!hbt]
\centering
\includegraphics[width=0.85\linewidth]{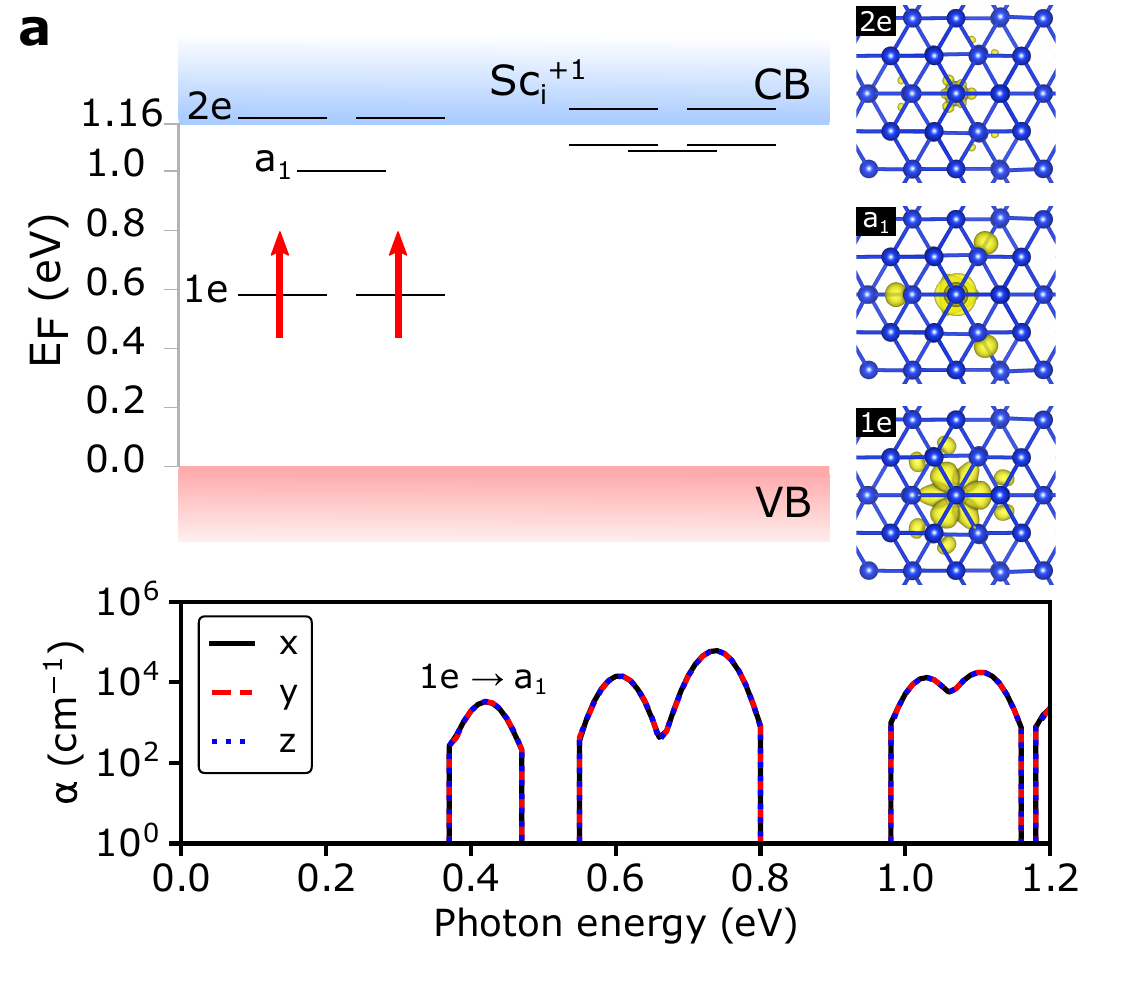}
\includegraphics[width=0.85\linewidth]{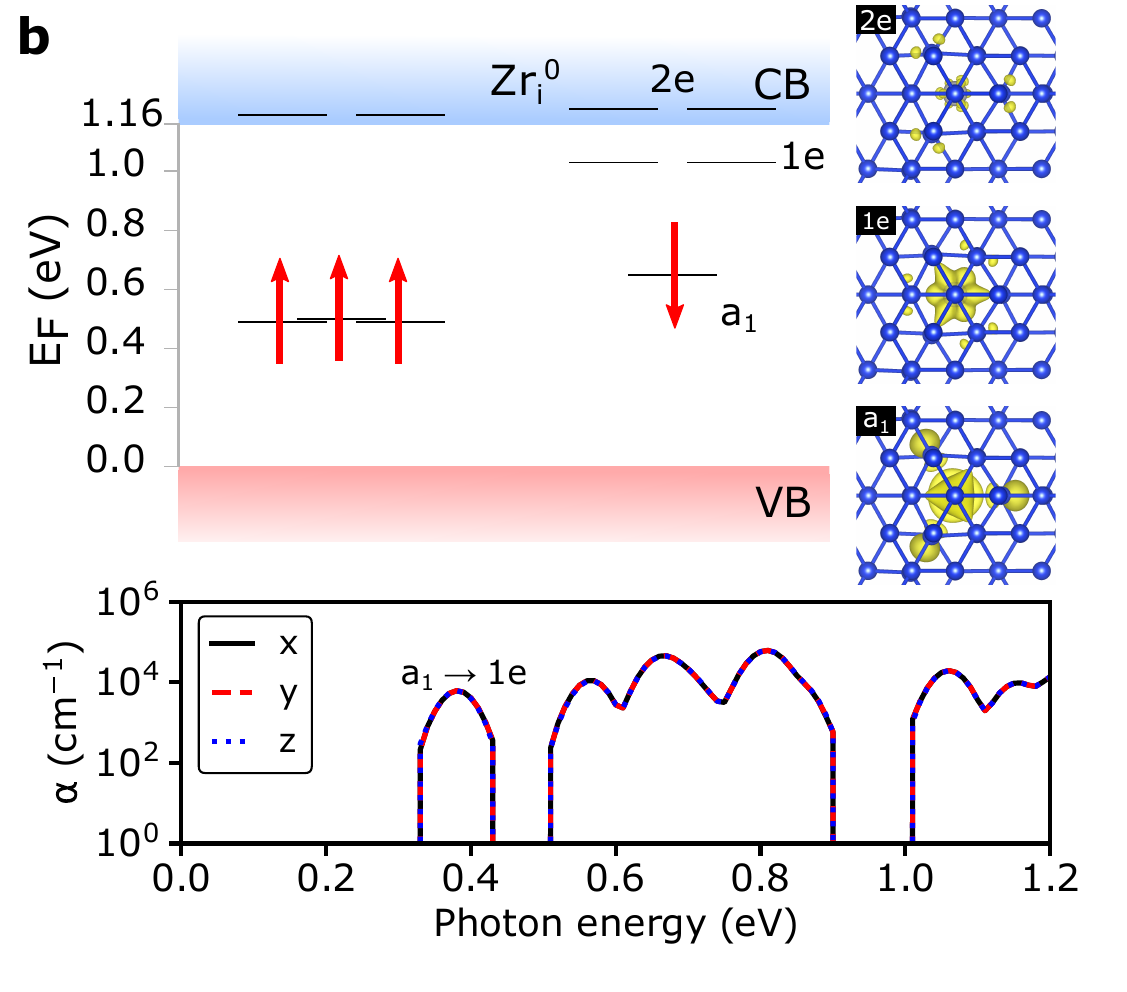}
\caption{\label{fig:int_candidate} 
\textbf{Single-particle defect level diagram and absorption spectra for representative interstitial defects of transition metals. }
Both (a) Sc$\mathrm{_{i}^{+1}}$ and (b) Zr$\mathrm{_{i}^{0}}$ have the C$_{3v}$ symmetry.
Numbers are used to distinguish different defect states while letters highlight their symmetries.
The isosurface plots of the charge density of corresponding defect states are shown at the density of 0.0025 $\frac{1}{a_{B}^{3}}$ and along the principle axis of the C$_{3}$ rotation, i.e., $<$111$>$.
Absorption coefficient ($\alpha$) are provided, with specific peaks indicated for the corresponding transitions between labeled defect states within the band gap.
x, y, and z indicate the polarization direction. 
All the other interstitial defect candidates (Zr$\mathrm{_{i}^{+2}}$ and V$\mathrm{_{i}^{+1}}$), which also have in-gap triplet-triplet optical transitions,  can be found in the Supplemental Materials.
}
\end{figure}

Interstitial sites are also very common for transition metal impurities in silicon. Two kinds of interstitials exist, those that have tetrahedral or hexagonal symmetry.
Here we focus only on interstitial defects at tetrahedral sites since, for transition metals, they generally have the lowest energies among all the interstitial defect sites \cite{Weber_APA_1983, Matsukawa_physb_2007}.
Applying the same selection procedure as substitutional defects, we first look at the CTL for interstitial defects as shown in Fig.\ \ref{fig:CTL_int}. 
We note that the correction scheme to address CTL with large E$\mathrm{_{NK}}$ is applied to Ti$\mathrm{_{i}}$, V$\mathrm{_{i}}$, Fe$\mathrm{_{i}}$, Co$\mathrm{_{i}}$, and Ni$\mathrm{_{i}}$ (see SM for details).
The criteria of at least two transition levels reduces the candidate list to Sc$\mathrm{_{i}}$, Ti$\mathrm{_{i}}$, V$\mathrm{_{i}}$, Mn$\mathrm{_{i}}$, Zn$\mathrm{_{i}}$, Zr$\mathrm{_{i}}$, and Au$\mathrm{_{i}}$.
After examining for spin triplet electronic configuration within the stable charge states, we drop Zn$\mathrm{_{i}}$ and Au$\mathrm{_{i}}$  out of the list due to their lack of the spin triplet configuration for Fermi energy within the band gap (see table S2 for the predicted total spin quantum numbers).
Mn$\mathrm{_{i}^{-1}}$, though has stable S=1 configurations, is dropped from the list since the associated defect levels are within the valence band.
A practical constraint of having one-electron levels sufficiently away from the band edges ($\Delta E$ $<$ 0.05 eV) further removes Ti$\mathrm{_{i}^{0}}$, Ti$\mathrm{_{i}^{+2}}$, and Sc$\mathrm{_{i}^{-1}}$ from the list. 
But we note that Ti$\mathrm{_{i}}$ can still be an plausible candidate when the generalized Koopmans' condition is \emph{fully} satisfied. 
Fig.\ \ref{fig:int_candidate} shows the single-particle defect level diagram, localized charge densities and the optical absorption spectra for two of the final four candidates, Sc$\mathrm{_{i}^{+1}}$ and Zr$\mathrm{_{i}^{0}}$ (see Fig.\ S3 for all the interstitial defect candidates).
The single-particle diagrams show that all of them have the $^{3}$A$_{2}$  triplet ground state, i.e., two unpaired electrons both on e orbitals (see Fig.\ \ref{fig:int_candidate} and Fig.\ S3).
Taking Sc$\mathrm{_{i}^{+1}}$ as example, the spin triplet excited states, $^{3}$E, forms when one of the up spin is excited from the e orbital to the a$_{1}$ orbital.
Similarly for Zr$\mathrm{_{i}^{0}}$,  $^{3}$E state forms when the down spin is promoted from a$_{1}$ to e orbitals. 
These interstitial defect systems closely resemble the NV center in diamond, which has C$_{3v}$ symmetry and $^{3}$A$_{2}$ - $^{3}$E transitions \cite{Doherty_NJP_2011}.
Charge densities show the C$_{3}$ symmetry along the principle axis, i.e., $<$111$>$,  consistent with the symmetry identified based on local defect configuration.
The absorption spectra show that the triplet-triplet optical transitions are allowed as shown by the first peaks. 
Similar to substitutional defects, the first peaks fall on the range between 0.4 to 0.6 eV and are far enough from other peaks.

In addition to substitutional and interstitial defects, we also investigate the Si-vacancy complexes, (Vac$\mathrm{_{Si}}$-TM$\mathrm{_{Si}}$),  for all the transition metals studied in this manuscript.
We find that there are no viable candidates with optical triplet-triplet transitions within the Si band gap, using the same workflow.
Specifically, they all share a similar property that the localized defect states are mostly within the bands, i.e., above the conduction and/or below the valence band edges of Si, preventing practical optical applications. 

\subsection*{Defect Formation Energetics}
\begin{figure}[!htb]
\centering
\includegraphics[width=0.95\linewidth]{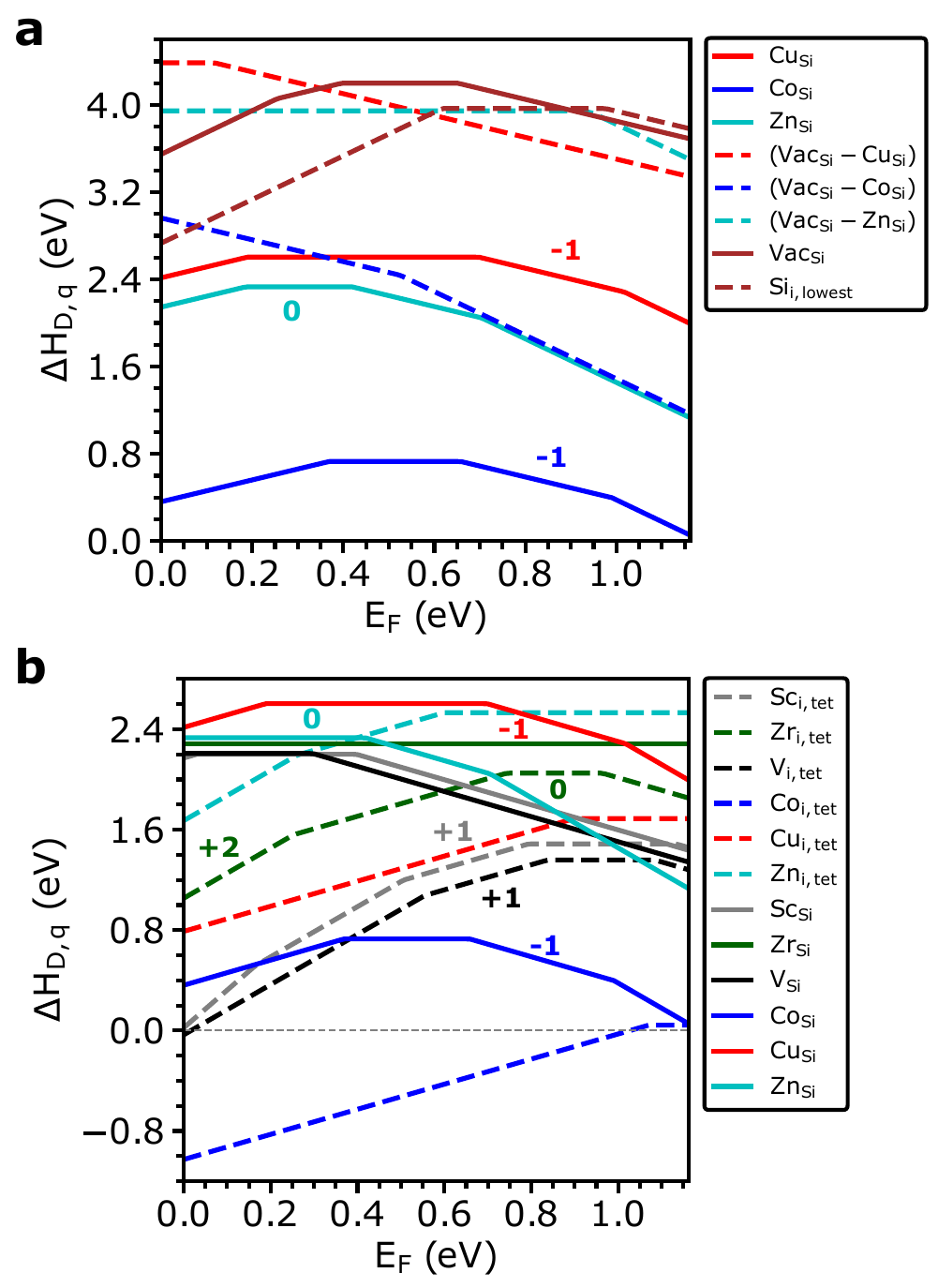}
\caption{\label{fig:vs_defect_complex} \textbf{Defect formation energy for candidate defects and their potential competing defects}
Comparison between (a) candidate substitutional defects and the vacancy complex counterparts and (b) candidate substitutional and interstitial defects are shown.
The number indicates the dominant charge state at a given Fermi energy range and highlights the charge states of the candidate substitutional and interstitial defects.
Note that the intermediate compounds are not considered here, i.e., total energies of the reference element states are used for the chemical potentials. 
Vac is used for Si vacancy to be distinct from Vanadium (V). 
For Si interstitials, only the one with the lowest energy among three defect types (tetrahedral, hexagonal, and split) at a given Fermi energy is shown.
}
\end{figure}


From the computational search, we identify seven TM impurities with specific combinations of charge states and defect types. 
Charge transition levels provide a guideline about how to achieve targeted defect charge states by tuning Fermi energy. 
This can be done via doping or electrical gating and the latter is preferred for semiconductor qubits to avoid charge or spin noises introduced by dopants. 
As a result, to maintain charge neutrality, the charged defects are mostly compensated by the electrons or holes supplied by the external circuit. 
At a given Fermi energy, in addition to the target TM impurities, the transition metal can occupy other defect sites.
These competing defects, along with Si vacancies and interstitials, can be charged and/or paramagnetic and act as charge and/or spin noises.
Furthermore, their relative abundances are important for the deterministic creation of the defects. 
Candidate defects that occupy the sites with the lowest defect formation energy are generally easier to create.  
In combination with the predicted total spin quantum numbers (see the full list in Supplemental Materials), we examine the defect energetics to address this concern. 

First, we compare the defect formation energies of the substitutional defects with their Si-vacancy-complex counterparts.  
Fig.\ \ref{fig:vs_defect_complex}(a) shows that for the candidate substitutional defects (indicated by their specific charge state), their competing silicon-vacancy-complex counterparts have significantly higher formation energies.
This suggests that the substitutional defects would have significantly higher defect concentration than their silicon-vacancy-complex counterparts.
In addition, Co$\mathrm{_{Si}^{-1}}$ will bind to Si vacancy if there is one nearby but the concentration of Si vacancy negligible due to large formation energy. 
Because of the focus on ion implantation as the means to produce the defects, Fig.\  \ref{fig:vs_defect_complex} neglects the effect of intermediate compounds on the chemical potential term shown in eq.\ \eqref{eq:DFE} but the relative position of the defect formation energies and thus the discussion above nonetheless remain the same. 
To be more specific, under the Si-rich condition, considering intermediate compounds will only change the chemical potential term for transition metals, which will change in the same way for both substitutional defects and their Si-vacancy-complex counterparts, leaving the difference in the formation energy unchanged.

Next, we compare defect formation energies of the substitutional defect candidates to the values of their interstitial counterparts and vice versa.
Fig.\ \ref{fig:vs_defect_complex}(b) shows the defect formation energies for candidate substitutional and interstitial transition metals in silicon (highlighted by their stable charge state).
For the three substituional defect candidates, Co$\mathrm{_{Si}^{-1}}$, Cu$\mathrm{_{Si}^{-1}}$, and Zn$\mathrm{_{Si}^{0}}$, 
within the Fermi energy range at which their relevant charge states are stable, we find that Cu$\mathrm{_{Si}^{-1}}$ and Co$\mathrm{_{Si}^{-1}}$ have higher formation energies than their interstitial counterparts.
This indicates, at equilibrium, they are harder to create compared to their  interstitial counterparts, but not necessarily rule them out from the candidate list. 
Unlike the rest, Zn$\mathrm{_{Si}^{0}}$ has similar formation energies to its interstitial counterpart, suggesting that it requires fine tune of Fermi level in order  to be the dominant defect. 
Similarly, we compare the defect formation energies of four interstitial defect candidates (Sc$\mathrm{_{i}^{+1}}$, Zr$\mathrm{_{i}^{0}}$, Zr$\mathrm{_{i}^{+2}}$, and V$\mathrm{_{i}^{+1}}$) to those of their substitutional counterparts.  
We find that they all have lower formation energies than their substitutional counterparts, suggesting less complexity of creating these candidate interstitial defects.

Lastly, we estimate the spin and/or charge noises that can be caused by their competing defects, along with native Si vacancies and interstitials. 
For the native defects, except for Vac$\mathrm{_{Si}^{+2}}$, Si$\mathrm{_{i}^{0}}$, and Si$\mathrm{_{i}^{+2}}$, they are mostly paramagnetic. 
But since their defect formation energies are much higher than those of candidate TM impurities, their defect concentrations will be negligible and thus the noises caused by the Si native defects are generally less concerning. 
At the Fermi energy range for Co$\mathrm{_{Si}^{-1}}$ to be the dominant charge state, Co$\mathrm{_{i}^{+1}}$ and $\mathrm{(Vac_{Si}-Co_{Si})^{-2}}$ are the competing defects.
Co$\mathrm{_{i}^{+1}}$ has zero spin but is charged while $\mathrm{(Vac_{Si}-Co_{Si})^{-2}}$  is both charged and paramagnetic. 
The interstitial has much lower formation energy than Co$\mathrm{_{Si}^{-1}}$, suggesting that the charge noise can be a major concern for physical realization. 
Similarly, Cu$\mathrm{_{i}^{0}}$, Cu$\mathrm{_{i}^{+1}}$and $\mathrm{(Vac_{Si}-Cu_{Si})^{-1}}$ are the competing defects for Cu$\mathrm{_{Si}^{-1}}$. 
Cu$\mathrm{_{i}^{0}}$ is charge neutral but paramagnetic while Cu$\mathrm{_{i}^{+1}}$ is charged but has zero spin.
Since Cu interstitial has much lower formation energy than Cu$\mathrm{_{Si}^{-1}}$, either charge or spin noise will be a concern, depending on the Fermi energy. 
$\mathrm{(Vac_{Si}-Cu_{Si})^{-1}}$ is  charged but has no spin. Its large formation energy gives rise to negligible concentration.
For Zn$\mathrm{_{Si}^{0}}$, pushing the Fermi energy close to the (+1/0) transition level is preferred to have Zn$\mathrm{_{i}^{+2}}$, which has zero spin as compared to the paramagnetic Zn$\mathrm{_{i}^{+1}}$.
However, since the difference in formation energy is small, charge noise by Zn$\mathrm{_{i}}$ is expected. 
On the other hand, $\mathrm{(Vac_{Si}-Zn_{Si})^{0}}$ is charge neutral and has no spin.

For Sc$\mathrm{_{i}^{+1}}$, its competing defects are both charged and $\mathrm{(Vac_{Si}-Sc_{Si})^{-2}}$ is also paramagnetic. 
But the noise is less concerning since both $\mathrm{(Vac_{Si}-Sc_{Si})^{-2}}$ and Sc$\mathrm{_{Si}^{-1}}$ have larger formation energy ($\gtrapprox$ 0.8 eV).
For V$\mathrm{_{i}^{+1}}$, its competing defects are all paramagnetic but the charge noise from $\mathrm{(Vac_{Si}-V_{Si})}$ and V$\mathrm{_{Si}}$ can be eliminated by pushing the Fermi energy toward VBM, which gives rise to zero charge state. 
This also increases the formation energy significantly and thus reduces the spin noise from the competing defects. 
Finally, for Zr$\mathrm{_{i}^{+2}}$ and Zr$\mathrm{_{i}^{0}}$, they share the same competing substitutional defect, Zr$\mathrm{_{Si}^{0}}$, which is both neutral and non-paramagnetic. 
$\mathrm{(Vac_{Si}-Zr_{Si})^{0}}$ and $\mathrm{(Vac_{Si}-Zr_{Si})^{-1}}$ are possible competing defects for Zr$\mathrm{_{i}^{+2}}$ and they are both paramagnetic.  
Tuning Fermi energy close to VBM can eliminate the charge noise by favoring neutral charge state and reduce the spin noise by increasing the formation energy. 
$\mathrm{(Vac_{Si}-Zr_{Si})^{-2}}$  is the competing defect for Zr$\mathrm{_{i}^{0}}$ and has zero spin. 
But due to its charge state and its similar formation energy to Zr$\mathrm{_{i}^{0}}$, charge noises caused by the silicon-vacancy complex can be detrimental to the lifetime of Zr$\mathrm{_{i}^{0}}$.
Based on the discussion above, interstitial defects are  better candidates than substitutional ones due to their generally lower formation energies than competing defects. 
%
\section*{\label{sec:discussion}Discussion}
%
\subsection*{Emerging Trends}
\begin{figure}[!htb]
\centering
\includegraphics[width=0.99\linewidth]{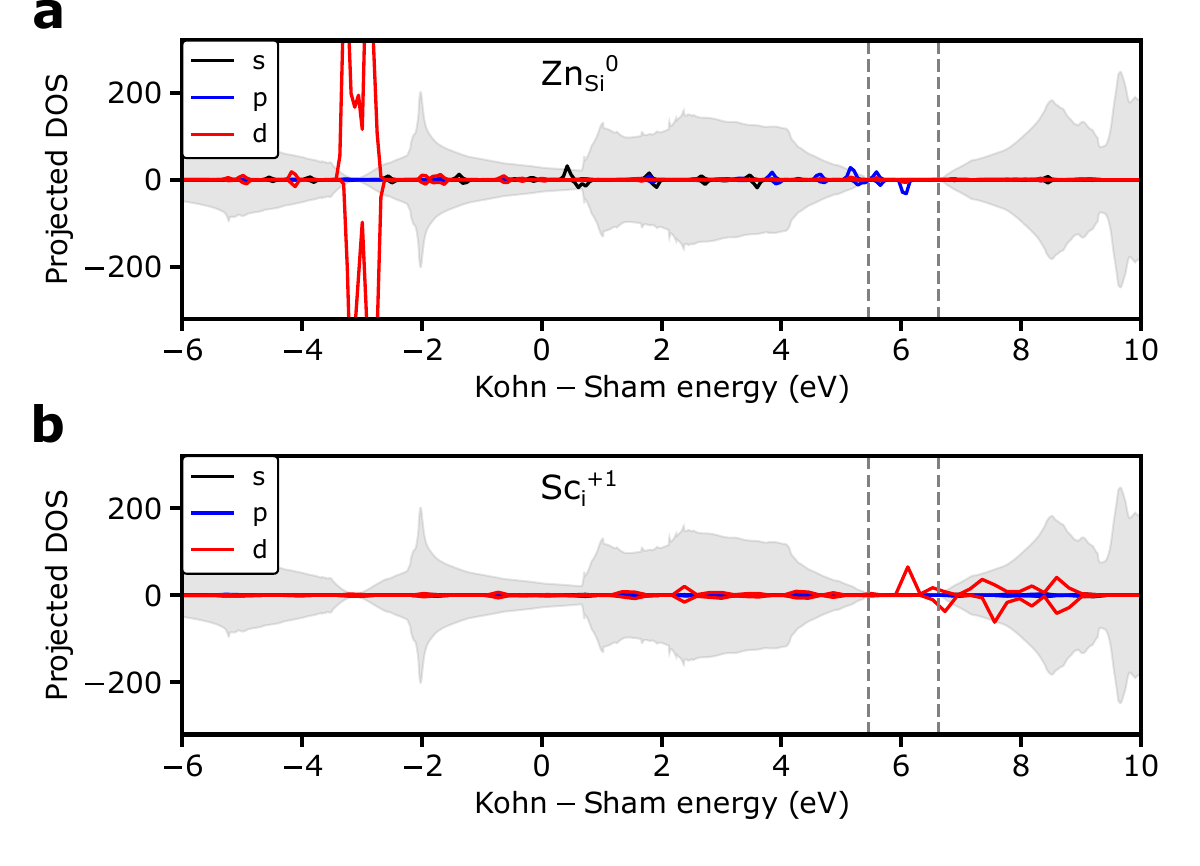}
\caption{\label{fig:pdos} \textbf{Projected density of state (PDOS) for transitional metal atoms.}
PDOS for transition metals in representative (a) substitutional  and (b) interstitial candidate defect systems are compared with density of state of bulk Si (shown by semi-transparent filled curve).
The magnitude of PDOS is amplified 40 times in order to emphasize the location.
Gray vertical dashed lines highlight the position of valence band maximum and conduction band minimum after potential alignment.
}
\end{figure}

Among the 3d and selected transition metals, we discover three substitutional and four interstitial defects in silicon, which have triplet-triplet optical transitions within the Si band gap.
As highlighted in Fig.\ \ref{fig:PT_trend}, we notice a general trend that these substitutional defect candidates are all located at the right side of the periodic table for transitional metals while these interstitial defect candidates are all located at the left side. 
To understand these trends, we first compare the electronic projected density of states (PDOS) on the atomic orbitals of the transition metal impurities to the density of states of bulk Si (see Fig.\ \ref{fig:pdos}, S6 and S7 for details).
This provides insights into the relative energies of the outermost $s$ and $d$ orbitals of the transition metals to the Si band structure.
We find out that, for substitutional defects, the outermost $s$ and $d$ orbitals of the transition metal are well below the valence band maximum of Si while, for interstitial defects, the outermost $s$ and $d$ orbitals are dispersed around the Si band gap.
Next, we count the number of participating electrons when the defects are created and use the relative energy positions identified previously to accommodate these electrons.
For substitutional defect candidates (see Fig.\ \ref{fig:pdos}(a) as example), there are sixteen electrons that need to be accommodated when a Zn$\mathrm{_{Si}^{0}}$ defect is created:
Zn has twelve valence electrons and the Si dangling bonds provide four electrons.
Given that outermost $s$ and $d$ orbitals of Zn (black and red curves in Fig.\ \ref{fig:pdos}(a), respectively) are deep below the valence band maximum, twelve of the sixteen electrons fill these states, with four remaining electrons available to occupy the defect states within the band gap.
As shown repetitively in Fig.\ \ref{fig:sub_candidate} and \ref{fig:int_candidate}, a triplet ground state requires either two or four electrons within the band gap.
For Zn$\mathrm{_{Si}^{0}}$, as shown in Fig.\ \ref{fig:sub_candidate}(b), we indeed have four electrons within the Si band gap, with three up spins and one down spin. 
Following the same procedure, Zn$\mathrm{_{Si}^{+2}}$ could be a plausible candidate but Fig.\ \ref{fig:CTL_sub} shows that the charge state of +2 is not stable within the band gap.
Similarly, all the transition metals at the Zn column could potentially work the same way if their outermost $s$ and $d$ orbital are deep below the valence band maximum.
For the transition metals in the columns near the Zn column, as long as the electron count meets the requirement, i.e., two or four electrons within the band gap, they could potentially have triplet ground states. 
However, having two or four electrons within the band gap can also result in spin singlet states, which are the cases for  Cu$\mathrm{_{Si}^{+1}}$ (see Fig.\ S4 for their defect-level diagrams). 
Nonetheless, this procedure explains the charge state of the substitutional defect candidates and their locations on the transition-metal periodic table. 
It could facilitate a quick search as long as the outermost $s$ and $d$ orbitals of transition metals are deep within the valence band maximum.  
Moving left-ward, the candidate list stop at Co column for silicon because, in order to fulfill the electron count, it requires more negative charge states, which are not stable within the Si band gap. 
But if a host material can accommodate more negative charge states and the relative position are similar, transition metals to the left of the Co column could also be substitutional defect candidates.

Similarly, electron counting also explains why interstitial defect candidates are all located at the left side of the transition-metal periodic table.
Unlike substitutional defects, the outermost $s$ and $d$ orbitals of the interstitial transition metals are located right near the band gap (see Fig.\ \ref{fig:pdos}(b) and S7).
Besides, unlike substitutional defects, no Si dangling bonds pointing to the defect site exist when an interstitial defect is created. 
Therefore, the total number of the outermost $s$ and $d$ electrons of a transition metal is the number of electrons that occupy the defect states within the silicon band gap. 
For instance, Sc has three outermost electrons, suggesting that Sc$\mathrm{_{i}^{+1}}$ has two electrons within the gap and could form a triplet ground state.
Moving rightward, the candidate list ends at the V column but it could potentially extend to Cr column with +2 charge state.
We ruled out Cr column for silicon because the triplet ground state of Cr$\mathrm{_{i}^{+2}}$  are localized slightly below the valence band maximum. 
Again, transition metals to the right of the V column are plausible if a different host semiconductor can accommodate more charge states. 
This is supported by the molybdenum color center observed in 4H and 6H-SiC \cite{Zhang_APR_2020}, which has significant larger band gap (E$_{g}$ $\approx$ 3.0-3.2 eV)\cite{Weber_PNAS_2010}.

Lastly, our findings are consistent with the observation that transition metals are portable among different host materials as long as they share the same lattice symmetry and similar bond length\cite{Diler_npj_QI_2020}.
Namely, when transition-metal color centers are identified in one host material, they are likely to be color centers in other \emph{similar} host materials as defined above.
We further identify the importance of the relative energy position of the outermost $s$ and $d$ orbitals with respect to the host band structure. 
A more comprehensive study is still required to test the concept of portable transition metals within the context of color centers. 

\subsection*{Significance and other in-gap transitions}
Currently, Si-based semiconductor qubits are limited in applications of quantum communication and sensing, due to the lack of established spin-photon interfaces and low operating temperature\cite{Chatterjee_arxiv_2020}, respectively. 
Aiming for NV-center-like defects in Si, our computational search finds transition metal impurities in silicon that also have optically allowed spin triplet-triplet transitions within the band gap, but with the wavelengths at the mid-IR range. 
Intuitively, these defect systems can potentially serve as the spin-photon interface, which couples and reads out spin donor qubits in Si, as an alternative scheme suggested by Morse \emph{et al.} \cite{Morse_SA_2017}.
Mid-IR range photons emitted by TM impurities have huge attenuation loss through the typical silica optical fibers, preventing its application for quantum communication at long distance, but they can be potentially used for free-space communication since their wavelengths fall on the atmospheric windows\cite{Hao_AO_2017,Lin_Nanophotonics_2018}.

In addition, the defect systems identified here also could have similar operation scheme to NV centers in diamond, i.e., optical initialization, allowing higher operating temperature for the application of quantum sensing. 
But the maximum temperature is expected to be lower than the value of the NV center for two reasons: 
First, small thermal excitation between the defect levels themselves and the conduction/valence band edges is needed to maintain the fidelity of the spin state\cite{Weber_PNAS_2010}. 
Generally speaking, host materials with larger band gap are expected to have larger energy difference between these states\cite{Weber_PNAS_2010}.
To compare the candidates identified here with the NV center, we estimate the number of thermally excited electrons using the smallest energy difference in the single-particle defect level diagram. 
The smallest energy difference falls in the range of 0.1-0.2 eV for TM impurities in Si while, in comparison, the reported values for the NV center are within the range of roughly 0.3-0.9 eV using HSE06\cite{Weber_PNAS_2010,Thiering_PRB_2018}. 
Using Boltzmann distribution with room temperature and energy difference of 0.15 and 0.6 eV, the number of thermally excited electrons are ~0.3 \% and ~0 \%, respectively. 
Based on this estimation,  we expect that the thermal excitation between these states is not the major limiting factor for TM impurities in Si, despite that they will lose some fidelity from thermally excited electrons.  
Secondly, higher operating temperature generally requires host materials with higher Debye temperature and smaller spin-orbit coupling in order to maintain a lifetime-limited optical line\cite{Weber_PNAS_2010, Wolfowicz_NRM_2021}. 
Since Si has a smaller Debye temperature and larger spin-orbit coupling than diamond, its maximum temperature is expected to be smaller than that of the NV center. 

Lastly, we prioritize the search for optical triplet-triplet transitions within the Si band gap as motivated by the underlying mechanism of the NV center in diamonds. 
However, TM impurities with singlet-singlet or doublet-doublet transitions also have the potential as spin-photon interfaces\cite{Lee_NNANO_2013,Hepp_PRL_2014}.
Since different operation schemes that utilize these two transitions are also possible, we also report all the identified transitions in the Supplementary Materials.
%

\subsection*{Further considerations}
%
Last but not the least, we emphasize that the computational search here is only the first step to screen candidate defect systems. 
In order to find an equivalent deep center in crystalline silicon that has the same (or similar) operation scheme as NV$^{-}$ centers in diamond, a few more steps are required to decide the final candidates.
Previous studies have shown the importance of having strong zero-field splitting (ZFS), which eliminates the need of external magnetic field to isolate the m=0 and m=$\pm 1$ states of a spin triplet system.\cite{Smart_arxiv_2020}
Therefore, future ZFS calculations\cite{Seo_PRMat_2017} can be used to further shorten the candidate list identified in this work.  

For the application of quantum communication, a sharp zero-phonon line (ZPL) of the optical spectra is required in order to create indistinguishable photons\cite{Chatterjee_arxiv_2020}. 
Since the rest of the spectra (phonon side bands) are not usable, a larger Debye-Waller factor (ratio between ZPL over phonon side bands) is desired.
Future calculations of ZPL and Debye-Waller factor\cite{Alkauskas_NJP_2014} can be used for comparisons with experimental spectra.

The way that intersystem crossing (ISC) of NV centers in diamond works enables the initialization of the qubit at room temperatures.
Based on this, another key next step toward Si-based qubit with higher operation temperature is to understand the ISCs in these candidates. 
ISC is the non-radiative relaxation of excited electron, mediated by phonon. 
With the development of first-principles phonon calculations in the past few decades\cite{Giustino_RMP_2017},
the ISC rate can now be quantitatively predicted and future calculations will finalize the candidate list.
 
Additionally, nuclear spins and diffusion coefficient of candidate transition metals are critical factors for physical realization of color centers in Si. 
Nuclear spins of defects generally act as spin noise, reducing the coherence time of the qubit system.
However, they can be also beneficial, serving as quantum memories. 
Therefore, selection criteria based on nuclear spins depend on the actual applications. 
Focused ion beam techniques can readily select the isotopes of a dopant and we notice that, among the candidate transition-metal defects, Sc, V, Co, and Cu have no naturally stable zero nuclear spin isotope.  
For the perspective of reliability, a mobile color center can undermine the performance of a Si-based qubit since all the external controls are exerted on predefined locations.
Mobility of a defect is measured by its diffusion coefficient and the values for transition metals in Si are well documented in the literature\cite{Weber_APA_1983,Graff_1995}.
Co and Cu are known to diffuse even at room temperature\cite{Weber_APA_1983}, making them less suitable if higher operating temperature is prioritized. 
V and Zn are also fairly mobile\cite{Graff_1995,Weber_APA_1983,Collins_APL_1966}, with diffusion coefficients at the order of 10$^{7}$ (cm$^{2}$/s) at temperature of 1100 $^{\circ}$C.
Among them, Zn has very low evaporation pressure, requiring encapsulation of the device\cite{Graff_1995}, while V stays in the bulk Si after cooling.
Lastly, Sc\cite{Zainabidinov_RPJ_2007} and Zr\cite{Sachdeva_PBCM_2006} have very low reported diffusion coefficient, making them essentially immobile after cooling. 
Based on both factors, Zr$\mathrm{_{i}^{0}}$ and Zr$\mathrm{_{i}^{+2}}$ are particularly promising for future experimental realizations. 

\section*{\label{sec:method}Methods}
\subsection*{Defect formation energy}
We followed the standard supercell approach to calculate the defect formation energies\cite{Lany_MSMSE_2009, Goyal_CMS_2017},
\begin{equation}\label{eq:DFE}
\Delta H_{D,q}(E_{F},\mu) =[E_{D,q}-E_{H}]+ \sum_{i} n_{i} \mu_{i}+qE_{F}+E_{corr}
\end{equation}
where $E_{D,q}$ and $E_{H}$ are generalized DFT total energies of the defect and host supercell, respectively. 
$\mu_{i}$ is the chemical potential of element $i$ and $n_{i}$ is the number of element $i$ added ($n_{i}$ $<$ 0) or removed ($n_{i}$ $>$ 0).
$q$ is the charge state of the defect and $E_{F}$ is the Fermi level (electron chemical potential) referenced to valence band maximum. 

$E_{corr}$ is the correction term that addresses the errors intrinsic to the supercell approach, including
(1) size effect (2) potential alignment (3) band edge problem.
Specifically, we follow the correction scheme devised by Lany and Zunger \cite{Lany_MSMSE_2009} and the implementation by Goyal \emph{et al.} \cite{Goyal_CMS_2017}.
To correct the image charge interaction for the 216-atom cell, static dielectric constant of 11.11 is used.
In this paper, the band gap problem is addressed via generalized hybrid density functional theory, specifically HSE06\cite{Krukau_JCP_2006}, using the standard mixing parameter value ($\alpha = 0.25$).
Such choice renders electronic indirect band gap of 1.16 eV, which is in great agreement with experimental values of  1.17 eV \cite{Persson_PRB_1996}.
HSE06 also predicts the lattice constant of 5.43 $\AA$, which agrees with experimental value \cite{Dargys_1994}.

Since the structural relaxation of the TM defect supercell can instead find a local minimum, we also check their competing spin configurations when the spin configuration is different from TM defect of the same group or from the neighboring TM of the same row. 
For example, we compare the total spin number of Au$_{Si}^{0}$ to Pt$_{Si}^{-1}$ and Cu$_{Si}^{0}$ since they are isoelectronic with each other in terms of valence electrons. 

\subsection*{Total energy calculations}
All the total energies of supercells are calculated using spin-polarized generalized hybrid density functional theory, specifically HSE06\cite{Krukau_JCP_2006} with the standard mixing parameter ($\alpha = 0.25$).
216-atom supercell with gamma-only kpoint sampling is used throughout the paper unless specified. 
We specifically use Vienna Ab initio Simulation Package (VASP 5.4.4 GPU version)\cite{Kresse_PRB_1996, Hacene_JCC_2012, Hutchinson_CPC_2012}.
The wavefunction is expanded using kinetic energy cutoff of 340 eV, which is sufficient for the projected augmented-wave (PAW) potentials\cite{Kresse_PRB_1999} used to describe the electron-ion interactions (the details for the PAW can be found in the Supplemental Materials).

\subsection*{Defect concentration}
Defect concentration is defined by,
\begin{equation}
C_{D,q} = C_{0}\exp (\frac{-\Delta H_{D,q}}{k_{B}T})
\end{equation}
where $C_{0}$ is the concentration of available sites, $\Delta H_{D,q}$ is the defect formation enthalpy and $k_{B}$ is the Boltzmann constant.
Here we assume negligible contribution from vibrational entropy. 

\subsection*{Thermodynamics charge transition level}
Thermodynamics charge transition level reveals the dominant charge state at a given Fermi level and is defined by the Fermi level that gives rise to the same defect formation energies for two charge states of $q_{1}$ and $q_{2}$,
\begin{equation}
(q_{1}/q_{2}) = \frac{E_{D,q_{1}}-E_{D,q_{2}}}{q_{2}-q_{1}}
\end{equation}
where $E_{D,q_{1(2)}}$ is the total energy of defect cell with charge of $q_{1}(q_{2})$ and the correction term for each defect cell is included.
Transition level of $(q_{1}/q_{2})$, like $E_{F}$ in Eq.\ \ref{eq:DFE}, is referenced to valence band maximum and it indicates that $q_{1}$ is the dominant charge state when the Fermi level is lower than the transition level.

\subsection*{Non-Koopmans' energy and correction}
\begin{figure}[!bht]
\centering
\hspace*{-0.2cm}\includegraphics[width=1.05\columnwidth]{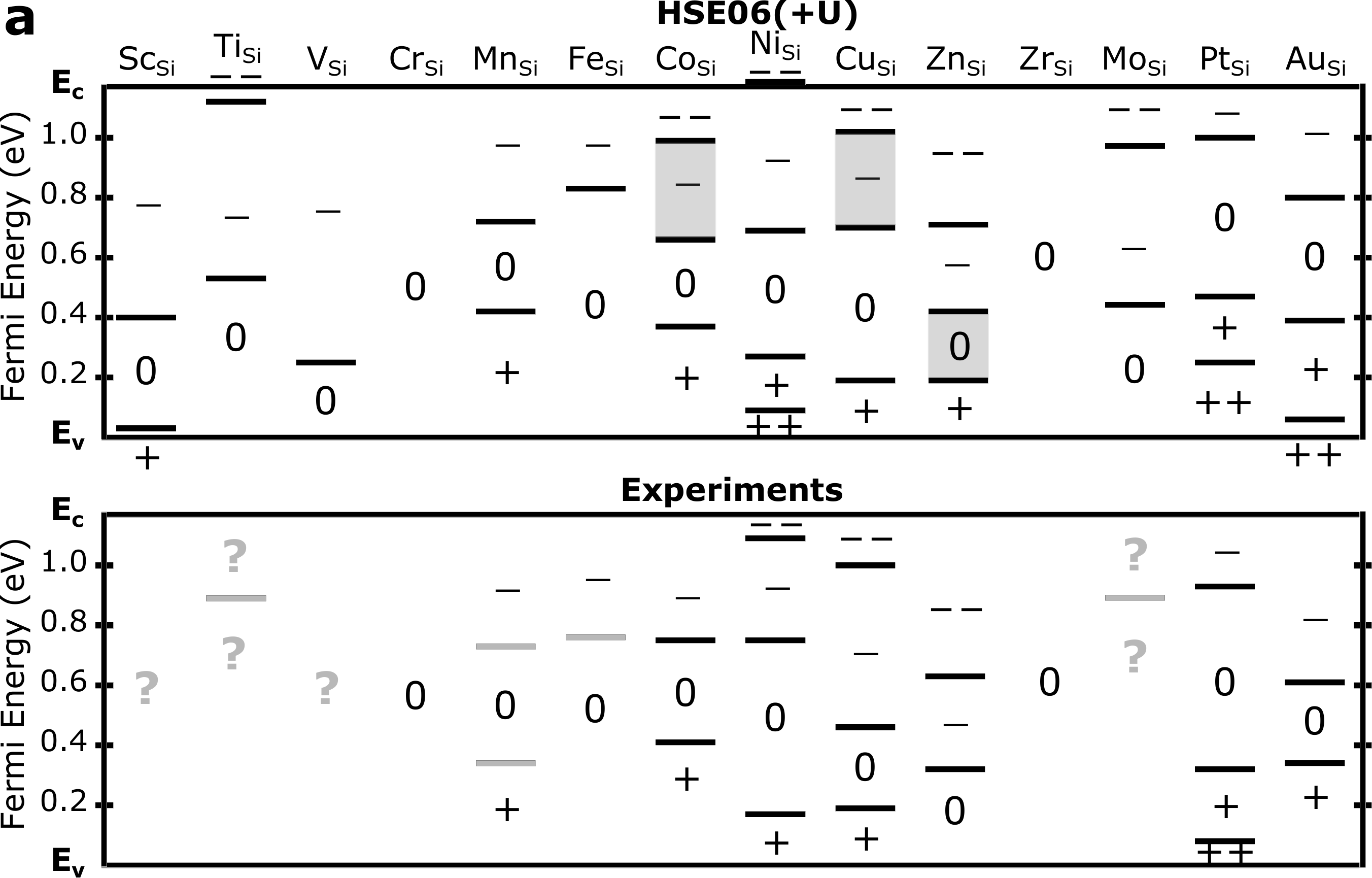}
\hspace*{-0.2cm}\includegraphics[width=1.05\columnwidth]{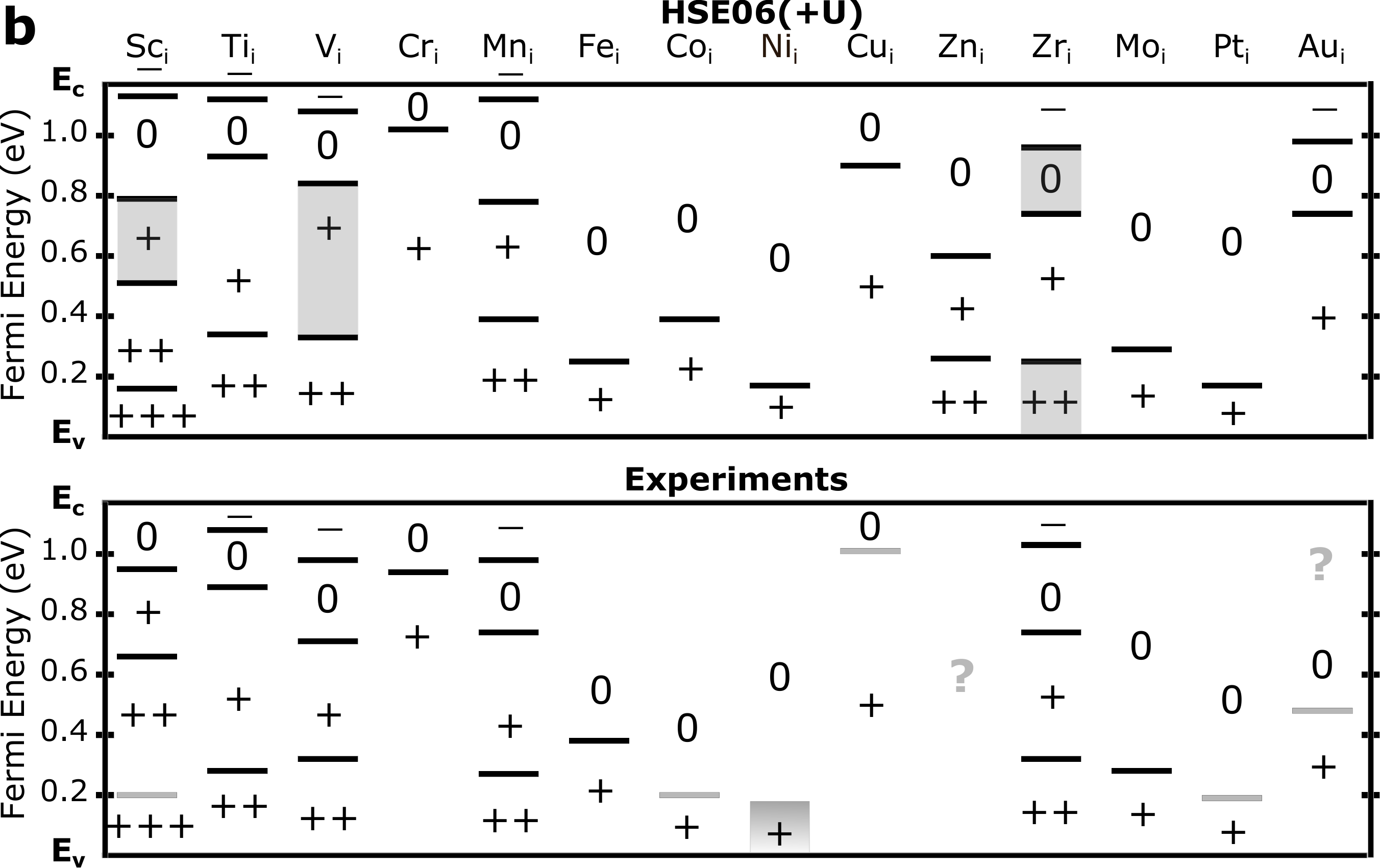}
\caption{\label{fig:CTL_compare} \textbf{Comparing predicted and experimental charge transition levels for 
(a) substitutional and (b) interstitial transition metal defects.}
Gray question marks and  horizontal bars indicate the charge transition levels and charge states are still inconclusive in the literature.
The gradient shade for interstitial Ni defect indicates that its CTL exists but the actual position is inconclusive.  "0" with no horizontal bars indicates that no electrically active defects are observed within the Si band gap.  $0$, $+$, and $-$ signs indicate the dominant charge state within the Fermi-level range. The gray shades highlight the candidates predicted in this work.
}
\end{figure}

Self-interaction error gives rise to nonlinear dependence of the total energy on the electron count\cite{Lany_PSSB_2011} and compliance with the generalized Koopmans'  condition ensures linearity and is formulated using the following equation
\begin{equation}
    E_{\mathrm{NK}} =  e_{N} - (E_{N}-E_{N-1}) 
\end{equation}
where N is the number of electrons, $E_{N}$ is the total energy for system with N electrons, and $e_{N}$ is the eigenvalue of the highest occupied state for the system with N electrons.
E$_{\mathrm{NK}}$ is the non-Koopmans' energy, which serves as an indicator for the deviation from the gKC.
E$_{\mathrm{NK}} > 0$ is commonly observed for semi-local functionals like GGA and is referred to as convex behavior 
\cite{Lany_PSSB_2011}. On the other hand, E$_{\mathrm{NK}} < 0$ is typical for Hartree-Fock method and is referred to as concave behavior. Enforcing the gKC (E$_{\mathrm{NK}}=0$) can solve the concave/convex problem and is generally done by either adjusting the U parameter in the DFT+U, or mixing parameter $\alpha$ for generalized hybrid DFT \cite{Lany_PSSB_2011}.  
For transition metals in Si, no single U or $\alpha$ can  guarantee to satisfy the gKC for both host material ($sp^{3}$ orbitals) and TM defects ($d$ orbitals).
To address this, we follow the approach by Iv\'ady \emph{et al.}\cite{Ivady_PRB_2013, Ivady_PRB_2014} and apply an occupation-dependent potential on $d$ orbitals of TM along with HSE06. 
HSE06 has been shown to comply with gKC for $sp^{3}$ semiconductor like Si\cite{Deak_PRB_2010} while the occupation-dependent potential only apply to  $d$ orbitals of TM. 
As a result, such approach can in principles satisfy gKC for both Si host and TM defects.
The occupation-dependent potential is equivalent to U term in  Dudarev’s implementation of the LDA + U method\cite{Dudarev_PRB_1988}, which is already implemented in VASP code.
Therefore, the correction is referred to as HSE06+U in this paper and the U is determined self-consistently to satisfy gKC. 
In practice, we ensure that $|$E$_{\mathrm{NK}}| < 0.1$ eV.

After applying the self-consistent correction to the charge transition levels with $|$E$_{\mathrm{NK}}| > $ 0.2 eV, Fig.\ \ref{fig:CTL_compare} shows the comparison between our HSE06(+U) and experimental results. 
Transition metal defects in Si are widely studied previously due to their critical roles as carrier traps. 
However, for the same reason, past studies focused only on transition metals and defect types that are relevant for carrier traps. 
As a results, only CTL of group 3-8 transition metals (Sc-Fe, Zr, Mo) interstitial defects and group 9-12 transition metals (Co-Zn,Pt,Au) substitutional defects are well-established in the literature.  
This is reflected by the gray question marks and horizontal bars in Fig.\ \ref{fig:CTL_compare}. 
For these well-established CTL, our HSE06(+U)  results generally agree very well  with experimental values considering the uncertainty of 0.1 eV in band edges and the remaining deviation from generalized Koopmans' condition (detailed comparison  in SM). 
The only two exceptions are the (-1/-2) level for Co$\mathrm{_{Si}}$ and (+1/0) level for Zn$\mathrm{_{Si}}$, to which no experimental deep-level transient spectroscopy peaks are assigned. 
Since these two CTL have low non-Koopmans' energy ( $|$E$_{\mathrm{NK}}| < $ 0.1 eV) and the energy difference from band edges are large ($>$0.1 eV),  they are highly likely to exist. 
But we note that they could prefer to form defect complex with common dopant and impurities in Si  like hydrogen, which are not considered in this work. 
Therefore, further experimental investigation are needed to resolve the difference. 

For those less studied TM defects Si, our HSE06(+U) results support some of the inconclusive assignments of CTL, like (+1/0) and (0/-1) levels for  Mn$\mathrm{_{Si}}$, (0/-1) level for Fe$\mathrm{_{Si}}$,(+2/+3) level for   Sc$\mathrm{_{i}}$and so on. 
Our results also challenge some of the  assignments like  (0/-1) level for Ti$\mathrm{_{Si}}$\cite{} and (0/-1) for Mo$\mathrm{_{Si}}$\cite{}. 
Lastly, our results predict some CTL that are either not experimentally observed or  assigned, like (+2/+1) and (+1/0) levels for Zn$\mathrm{_{i}}$ and (0/-1) level for Sc$\mathrm{_{Si}}$. 
In short, in addition to the goal of identifying promising optically active defects, our HSE06(+U) results for charge transition levels fill some of the gaps in the literature for transition metal defects in Si.

\subsection*{Selection criterion for charge transition level}
Charge transition levels (CTL) within band gap are useful as they determine the number of electrons present in mid-gap defect states as a function of the Fermi energy and are thus helpful in revealing the potential triplet-triplet optical transitions within the Si band gap.
For instance, zero CTL indicates shallow defects when the effective hydrogenic state is not included. 
On the other hand, one or more CTL suggest localized defects and one or more electron states within the band gap, respectively.
We also note that, depending on the actual system,  one CTL such as (+2/0) can correspond to discharging of two electrons at the same time when the Fermi energy is shifted below the CTL.  
As a result, the number CTL can be smaller than the number of electron states. 

For a given impurity defect in Si to accommodate both a spin triplet electronic ground state and a spin triplet excited state within the band gap of the host material, at least three electron states within the band gap are needed (see Fig.\ \ref{fig:sub_candidate}(a) or \ref{fig:int_candidate}(a) as example).
For a spin triplet ground state to exist, two electron states are needed and this corresponds to one or two CTL, depending on the symmetry of the defect and thus on the degeneracy of the defect electron states. 
Therefore, there must be at least one CTL that corresponds to discharging of the spin triplet ground state.
Similarly, for a spin triplet excited state to form a spin triplet ground state, at least another empty electron state within the band gap is needed. Thus, at least one more CTL is needed and together we need at least two CTL to accommodate triplet-triplet transition within the band gap.

\subsection*{Single-particle defect levels}
Single-particle defect levels are the single-particle Kohn-Sham orbitals of a defect cell and potential alignment is performed to align them with the band structure of the host cell. 
We note that single-particle states within the electronic band gap correlate with the thermodynamics charge transition levels but by no means the same. 

\subsection*{Optical absorption coefficient}
Absorption coefficient ($\alpha$) at a given energy difference($\epsilon_{i}$)  is calculated using the complex refractive index (n),
\begin{equation}
\alpha (\epsilon_{i})=\frac{2\epsilon_{i} \operatorname{Im}[n(\epsilon_{i})]}{\hbar c}
\end{equation}
where $\hbar$ is the reduced Planck's constant, c is the speed of light.
The complex refractive index is the square root of the frequency-dependent complex dielectric function, in which the imaginary part can be calculated using the linear response approach and the real part by Kramers-Kronig transformation\cite{Gajdos_PRB_2006}. 
We use the VASP code (VASP 5.4.4) and the linear response calculation is based on the generalized Kohn-Sham states of HSE06 calculation for candidate defect systems.
The complex shift of 0.009 is used in the Kramers-Kronig transformation and the energy resolution is 0.01 eV.
Around 800 empty bands are used for the supercell with around 864 electrons, which are sufficient for the small photon energy range of interest since the difference in absorption intensity is negligible with increasing empty bands.

\section*{Data availability}
The data are available upon reasonable request. 

\begin{acknowledgments}
This work was funded by NREL’s Laboratory Directed Research and Development program. NREL is supported by the US Department of Energy under Contract No. DE-AC36-08GO28308 with Alliance for Sustainable Energy, LLC, the Manager and Operator of the National Renewable Energy Laboratory. 
The research was performed using computational resources sponsored by the Department of Energy's Office of Energy Efficiency and Renewable Energy and located at the National Renewable Energy Laboratory.
\end{acknowledgments}

\section*{Author Contributions}
C.-W. L. performed the defect calculations and analyzed the results.
V.S. and C.-W.L. designed the simulations. 
V.S. oversaw the calculations and the analysis. 
V.S.,  M.S., and A.T.  conceived the idea. 
All authors contributed to the writing of the manuscript.

\section*{Competing Interests statement}
The authors declare no competing	 interests. 
\bibliography{Si_qubit.bib}
\end{document}